\documentstyle[aps]{revtex}

\begin{document}
\draft
\title{Small-World Network Effect in Competing Glauber- and Kawasaki-type Dynamics}
\author{Jian-Yang Zhu$^{1,2,}\thanks{%
Email address: zhujy@bnu.edu.cn}$,Wei Liu$^2$, Han Zhu$^3$}
\address{$^1$CCAST (World Laboratory), Box 8730, Beijing 100080, China.\\
$^2$Department of Physics, Beijing Normal University, Beijing 100875, China.%
\thanks{%
Mailing address}\\
$^2$Department of Physics, Nanjing University, Nanjing 210093, China.}
\maketitle

\begin{abstract}
In this article, we investigate the competing Glauber-type and Kawasaki-type
dynamics with small-world network (SWN) effect, in the framework of the
Gaussian model. The Glauber-type single-spin transition mechanism with
probability $p$ simulates the contact of the system with a heat bath and the
Kawasaki-type dynamics with probability $1-p$ simulates an external energy
flux. Two different types of SWN effect are studied, one with the total
number of links increased and the other with it conserved. The competition
of the dynamics leads to an interesting self-organization process that can
be characterized by a phase diagram with two identifiable temperatures. By
studying the modification of the phase diagrams, the SWN effect on the two
dynamics is analyzed. For the Glauber-type dynamics, more important is the
altered average coordination number while the Kawasaki-type dynamics is
enhanced by the long range spin interaction and redistribution.

\end{abstract}


\section{Introduction}

Many systems in nature or society can be well described by small world
networks (SWN), which were first proposed by Watts and Strogatz in Ref. \cite
{Watts}. SWNs are those intermediate between a regular lattice and a random
network. They can be realized by introducing a very small portion of long
range links to a regular lattice. Such networks appear as small-worlds like
random graphs, i.e., with a small average node-node distance that grows
logarithmically with the network size. Meanwhile they also have usually
large clustering coefficients \cite{nature-review,Albert,newman-review}.
Since the first model system was proposed\cite{Watts}, a large literature
has appeared and the properties of various models and processes on SWNs have
been extensively studied, including percolation, coloring, coupled
oscillators, iterated games, diffusion processes, epidemic processes, and
spin models (see Ref. \cite{nature-review,Albert,newman-review} and
references therein). For example, it has been recently found that some
spin-lattice models belonging to different universality classes now show
mean-field behavior on SWNs\cite
{newman-review,Ising1,Ising2,Ising3,Ising4,XY}. Furthermore, recently Zhu 
{\it et al.} introduced SWN effect to critical dynamics \cite{zhu}, and the
investigation has been extended to the kinetic properties of the spin models.

According to Glauber's and Kawasaki's theory \cite{Glauber,Kawasaki}, the
time evolution of the order parameters can be described by a Markov process
with a specific choice of the dynamic mechanism. Two mechanisms have been
extensively studied: Glauber's single-spin flip mechanism\cite{Glauber} with
the order parameter nonconserved and Kawasaki's spin-pair exchange mechanism%
\cite{Kawasaki} with the order parameter conserved. In the past decade, an
interesting problem has been attracting much attention, i.e., the competing
Glauber-type and Kawasaki-type dynamics. We briefly explain the physical
ground: Glauber's mechanism with probability $p$ is used to simulate the
contact of the system with a heat bath and favors a low energy state.
Meanwhile Kawasaki's mechanism with probability $1-p$ simulates an external
energy flux and, naturally, favors a high energy state. Then there will a
competition between the two dynamics. This competing mechanism has been
applied to the spin models\cite{com1,com2,com3,com4,com5,com6,zhu3}, and
interesting self-organization has been reported. All these works have
considered regular lattice, and to extend the investigation to the
small-world networks should be of interest. At the same time, such a study
may also have theoretical meaning. In Zhu's work\cite{zhu}, it has been
found that the SWN effect may have different types of influence on the
Glauber-type dynamics and the Kawasaki-type dynamics. By focusing on the
competition of the dynamics, it is easy for us to further understand the
influence of the SWN effect and highlight the disparities between the
dynamics.

In this article we investigate the SWN effect on the competing dynamics, and
we demonstrate the conclusions by comparing the phase diagrams obtained with
and without the SWN effect. Before we present the calculations, we first
describe the prototypes of SWNs that are used in the present study, and then
briefly review the generalized competing mechanism.

{\it The SWN models}: After the first prototype of SWN was proposed by Watts
and Strogatz\cite{Watts}, there appear a variety of model systems that can
be categorized into the two groups, i.e., with the total number of links
increased or conserved. Correspondingly, in the present research we study
two specific, while representative models described below: (1) in a
one-dimensional loop, for example, each randomly selected pair of vertices
are additionally connected with probability $p_A$; (2) the vertices are
visited one after another, and each of its links in the clockwise sense is
left in place with probability $1-p_R$ and is reconnected to a randomly
selected other node with probability $p_R$. Networks of higher dimensions
can be similarly built. We call the first model adding-type small-world
network (A-SWN) and the second one rewiring-type network (R-SWN).

{\it The generalized competing mechanism}: Glauber's flip mechanism and
Kawasaki's exchange mechanism were originally presented for the Ising model.
Recently they have been generalized to single-spin transition mechanism\cite
{zhu1} and spin-pair redistribution mechanism\cite{zhu2}, respectively,
which can be applied to arbitrary spin systems. Correspondingly there is the
generalized competing dynamics\cite{zhu3}, which provides a basis for the
discussions in this article. With the competing mechanism, the master
equation can be written as 
\begin{equation}
\frac{d}{dt}P\left( \{\sigma \},t\right) =pG_{me}+\left( 1-p\right) K_{me},
\label{1}
\end{equation}
where $pG_{me}$ denotes the single-spin transition with probability $p$ and $%
\left( 1-p\right) K_{me}$ denotes the spin-pair redistribution with
probability $1-p$. For your convenience we list in Sec. II and III some
important equations of the transition mechanism and the redistribution
mechanism, the details of which can be found in respective references.

In this article we study the generalized competing dynamics in the framework
of the kinetic Gaussian model, which is mathematically tractable. It has
been revealed in Ref. \cite{zhu} that the SWN effect on the Gaussian model
is of the mean-field nature. Considering this, in the present study a
simplified method is adopted, although a rigorous treatment is possible.

In Sec. \ref{Sec.2} we study the competing mechanics on A-SWN and In Sec. 
\ref{Sec.3}, we extend the discussions to R-SWN. Section \ref{Sec.4} is the
summarization with some discussions.

\section{Competing mechanism on A-SWN}

\label{Sec.2}

In this section we present our discussion of the competing dynamics on
adding-type small-world networks (the definition see the Introduction). In
subsection \ref{Subsec.2.1} through \ref{Subsec.2.3} we give the formulation
of the Glauber-type mechanism, the Kawasaki-type mechanism and the competing
mechanism, and then in subsection \ref{Subsec.2.4} the competing mechanism
is applied to the three-dimensional Gaussian model.

\subsection{The Glauber-type mechanism}

\label{Subsec.2.1}

With the Glauber-type single-spin transition mechanism\cite{zhu,zhu1} on
A-SWN, the master equation can be written as 
\begin{equation}
\frac{d}{dt}P(\{\sigma \},t)=-\sum\limits_{i}\sum\limits_{\sigma }\left[
W_{i}\left( \sigma _{i}\rightarrow \hat{\sigma}_{i}\right) P\left( \{\sigma
\},t\right) -W_{i}\left( \hat{\sigma}_{i}\rightarrow \sigma _{i}\right)
P\left( \{\sigma _{j\neq i}\},\hat{\sigma},t\right) \right] ,  \label{3}
\end{equation}
where the transition probability 
\[
W_{i}\left( \sigma _{i}\rightarrow \hat{\sigma}_{i}\right) =\frac{1}{Q_{i}}%
\exp \left[ -\beta {\cal H}\left( \left\{ \sigma _{j\neq i}\right\} ,\hat{%
\sigma}_{i}\right) \right] ,Q_{i}=\sum\limits_{\hat{\sigma}_{i}}\exp \left[
-\beta {\cal H}\left( \left\{ \sigma _{j\neq i}\right\} ,\hat{\sigma}%
_{i}\right) \right] . 
\]
With equation (\ref{3}), one can continue to write that\cite{zhu,zhu1}, 
\begin{equation}
\frac{d}{dt}q_{k}\left( t\right) =Q_{k}^{G}=-q_{k}\left( t\right)
+\sum\limits_{\{\sigma \}}\left[ \sum\limits_{\hat{\sigma}_{k}}\hat{\sigma}%
_{k}W_{k}\left( \sigma _{k}\rightarrow \hat{\sigma}_{k}\right) \right]
P\left( \{\sigma \},t\right) ,  \label{q-Glauber}
\end{equation}
where 
\begin{equation}
q_{k}\left( t\right) =\sum\limits_{\left\{ \sigma \right\} }\sigma
_{k}P\left( \{\sigma \},t\right) .  \label{4}
\end{equation}

\subsection{The Kawasaki-type mechanism}

\label{Subsec.2.2}

With the Kawasaki-type spin-pair redistribution mechanism on A-SWN, the
master equation can be written as\cite{zhu,zhu2} 
\begin{eqnarray*}
\frac{d}{dt}P(\{\sigma \},t) &=&\sum\limits_{\langle jl\rangle }\sum\limits_{%
\hat{\sigma}_{j}\hat{\sigma}_{l}}\left[ -W_{jl}(\sigma _{j}\sigma
_{l}\rightarrow \hat{\sigma}_{j}\hat{\sigma}_{l})P(\{\sigma \},t)+W_{jl}(%
\hat{\sigma}_{j}\hat{\sigma}_{l}\rightarrow \sigma _{j}\sigma
_{l})P(\{\sigma _{j\neq i},\sigma _{l\neq k}\},\hat{\sigma}_{j},\hat{\sigma}%
_{l},t)\right] \\
&&+\frac{1}{2}p_{A}\sum\limits_{j}\sum\limits_{l\neq j}\sum\limits_{\hat{%
\sigma}_{j}\hat{\sigma}_{l}}\left[ -W_{jl}(\sigma _{j}\sigma _{l}\rightarrow 
\hat{\sigma}_{j}\hat{\sigma}_{l})P(\{\sigma \},t)\right. \\
&&\left. +W_{jl}(\hat{\sigma}_{j}\hat{\sigma}_{l}\rightarrow \sigma
_{j}\sigma _{l})P(\{\sigma _{j\neq i},\sigma _{l\neq k}\},\hat{\sigma}_{j},%
\hat{\sigma}_{l},t)\right] ,
\end{eqnarray*}
where the redistribution mechanism 
\begin{equation}
W_{jl}(\sigma _{j}\sigma _{l}\rightarrow \hat{\sigma}_{j}\hat{\sigma}_{l})=%
\frac{1}{Q_{jl}}\delta _{\sigma _{j}+\sigma _{l},\hat{\sigma}_{j}+\hat{\sigma%
}_{l}}\exp [-\beta H(\{\sigma _{m}\}_{m\neq j,l},\hat{\sigma}_{j},\hat{\sigma%
}_{l})],  \label{5}
\end{equation}
\[
Q_{jl}=\sum\limits_{\hat{\sigma}_{j}\hat{\sigma}_{l}}\delta _{\sigma
_{j}+\sigma _{l},\hat{\sigma}_{j}+\hat{\sigma}_{l}}\exp [-\beta H(\{\sigma
_{m}\}_{m\neq j,l},\hat{\sigma}_{j},\hat{\sigma}_{l})]. 
\]
With the master equation above, we can further get that 
\begin{eqnarray}
\frac{d}{dt}q_{k}(t) &=&Q_{k}^{K}=-2Dq_{k}(t)+\sum\limits_{\{\sigma
\}}\sum_{\omega }\left[ \sum\limits_{\hat{\sigma}_{k},\hat{\sigma}_{k+\omega
}}\hat{\sigma}_{k}W_{k,k+\omega }\left( \sigma _{k}\sigma _{k+\omega
}\rightarrow \hat{\sigma}_{k}\hat{\sigma}_{k+\omega }\right) \right]
P(\{\sigma \},t)  \nonumber \\
&&+p_{A}\{-(N-1)q_{k}(t)+\sum\limits_{\{\sigma \}}\left[ \sum\limits_{l\neq
k}\sum\limits_{\hat{\sigma}_{k},\hat{\sigma}_{l}}\hat{\sigma}%
_{k}W_{kl}\left( \sigma _{k}\sigma _{l}\rightarrow \hat{\sigma}_{k}\hat{%
\sigma}_{l}\right) \right] P(\{\sigma \},t)\},  \label{q-Kawasaki}
\end{eqnarray}
where $\sum_{\omega }$ denotes the summation taken over nearest neighbors.

\subsection{Competing mechanism}

\label{Subsec.2.3}

Naturally, both Glauber's dynamics and Kawasaki's dynamics favor a lower
energy state. However, when the system under study is in contact with a heat
bath while exposed to an external energy flux, one requires a competition
between a process that favors lower system energy and another process that
favors higher system energy. Usually, Glauber's mechanism is used to
simulate the contact of the system with a heat bath and favors a lower
energy state. Meanwhile Kawasaki's mechanism can be modified in order to
simulate an external energy flux that drives the system towards higher
energy. This can be achieved by switching $\beta $ to $-\beta $, or $K=\beta
J=J/K_BT$ to $-K$, and modifying the redistribution probability accordingly,
i.e. 
\begin{equation}
W_{jl}\propto \exp [-\beta H]\Rightarrow W_{jl}\propto \exp [+\beta H]
\end{equation}
This means that the competition between Glauber's mechanism and Kawasaki's
mechanism is actually that between ferromagnetism and antiferromagnetism.

Based on the above considerations, we use the competing mechanism to
simulate the dynamics of a system in contact with a heat bath and exposed to
an external energy flux simultaneously. The master equation can be written
as (Eq. (\ref{1})), 
\[
\frac{d}{dt}P\left( \{\sigma \},t\right) =pG_{me}+\left( 1-p\right) K_{me}, 
\]
where 
\[
G_{me}=-\sum\limits_{i}\sum\limits_{\sigma }[W_{i}(\sigma _{i}\rightarrow 
\hat{\sigma}_{i})P(\{\sigma \},t)-W_{i}(\hat{\sigma}_{i}\rightarrow \sigma
_{i})P(\{\sigma _{j\neq i}\},\hat{\sigma},t). 
\]
and 
\begin{eqnarray*}
K_{me} &=&\sum\limits_{\langle jl\rangle }\sum\limits_{\hat{\sigma}_{j}\hat{%
\sigma}_{l}}\left[ -W_{jl}(\sigma _{j}\sigma _{l}\rightarrow \hat{\sigma}_{j}%
\hat{\sigma}_{l})P(\{\sigma \},t)\right. \\
&&\left. +W_{jl}(\hat{\sigma}_{j}\hat{\sigma}_{l}\rightarrow \sigma
_{j}\sigma _{l})P(\{\sigma _{j\neq i},\sigma _{j\neq i}\},\hat{\sigma}_{j,}%
\hat{\sigma}_{l},t)\right] \\
&&+\frac{1}{2}p_{A}\sum\limits_{j}\sum\limits_{l\neq j}\sum\limits_{\hat{%
\sigma}_{j}\hat{\sigma}_{l}}\left[ -W_{jl}(\sigma _{j}\sigma _{l}\rightarrow 
\hat{\sigma}_{j}\hat{\sigma}_{l})P(\{\sigma \},t)\right. \\
&&\left. +W_{jl}(\hat{\sigma}_{j}\hat{\sigma}_{l}\rightarrow \sigma
_{j}\sigma _{l})P(\{\sigma _{j\neq i},\sigma _{l\neq k}\},\hat{\sigma}_{j,}%
\hat{\sigma}_{l},t)\right] .
\end{eqnarray*}
From the master equation, we can obtain the evolving equation of single
spins, 
\begin{equation}
\frac{d}{dt}q_{k}(t)=pQ_{k}^{G}+(1-p)Q_{k}^{K}.  \label{6}
\end{equation}
The first term describes the Glauber-type dynamics which is used to simulate
the influence of the energy flux and $Q_{k}^{G}$ is given by Eq. (\ref
{q-Glauber}). The second term describes the Kawasaki-type dynamics which is
used to simulate the influence of the energy flux. As mentioned above, we
require it to favor higher energy state. However, the redistribution
probability, Eq. (\ref{5}), clearly favors a lower energy state. Considering
this, we can use the expression of Eq. (\ref{q-Kawasaki}) for $Q_{k}^{K}$,
if we simply switch the sign before $\beta $ in the redistribution
probability.

\subsection{The competing dynamics in the Gaussian model}

\label{Subsec.2.4}

Now we turn to the Gaussian model built on an A-SWN, of which the
Hamiltonian can be written as 
\begin{eqnarray}
-\beta {\cal H}\left( \left\{ \sigma \right\} \right)
&=&K\sum\limits_{ijk}\sigma _{ijk}\left( \sigma _{i+1,j,k}+\sigma
_{i,j+1,k}+\sigma _{i,j,k+1}\right)  \nonumber \\
&&+\frac 12p_AK\sum\limits_{ijk}\sigma _{ijk}\sum\limits_{i^{\prime
}j^{\prime }k^{\prime }}\sigma _{i^{\prime }j^{\prime }k^{\prime }}-\frac 12%
p_AK\sum\limits_{ijk}\sigma _{ijk}^2,  \label{A-SWN}
\end{eqnarray}
where $\beta =1/k_BT$ and $K=\beta J$. We only discuss the case of $K>0$,
namely, $J>0$. This case corresponds to the ferromagnetic system. The
expression (\ref{A-SWN}) is obtained by taking into consideration the fact
that the influence of the system as a whole on individual spins is of the
mean-field nature\cite{zhu} (for related discussions see Ref. \cite
{newman-review,Ising1,Ising2,Ising3,Ising4,XY}). The spins can take any real
value from $-\infty $ to $+\infty $. The probability of finding a given spin
between $\sigma _k$ and $\sigma _k+d\sigma _k$ is assumed to be the
Gaussian-type distribution, $f(\sigma _k)$ $d\sigma _k=\sqrt{\frac b{2\pi }}%
\exp \left( -\frac b2\sigma _k^2\right) d\sigma _k$, where $b$ is a
distribution constant independent of temperature. Therefore, the summation
of the spin value turns into an integration, and we can further obtain from
Eqs. (\ref{q-Glauber}) and (\ref{q-Kawasaki}), 
\[
Q_{ijk}^G=-q_{ijk}(t)+\frac Kb\sum\limits_{\omega =\pm 1}(q_{i+\omega
,j,k}(t)+q_{i,j+\omega ,k}(t)+q_{i,j,k+\omega }(t))+\frac Kb(N-1)p_AM(t),\
(K>0) 
\]
where $M\left( t\right) \equiv \frac 1N\sum_kq_k\left( t\right) $, and 
\begin{eqnarray}
Q_{ijk}^K &=&\frac 1{2\left[ b+(-K)\right] }b\left\{ \left[
(q_{i+1,j,k}-q_{i,j,k})-(q_{i,j,k}-q_{i-1,j,k})\right] \right.  \nonumber \\
&&\left. +\left[ (q_{i,j+1,k}-q_{i,j,k})-(q_{i,j,k}-q_{i,j-1,k})\right]
+\left[ (q_{i,j,k+1}-q_{i,j,k})-(q_{i,j,k}-q_{i,j,k-1})\right] \right\} 
\nonumber \\
&&+\frac{\left( -K\right) }{2\left[ b+(-K)\right] }\left[
2(2q_{i-1,j,k}-q_{i-1,j+1,k}-q_{i-1,j-1,k})+(2q_{i-1,j,k}-q_{i,j,k}-q_{i-2,j,k})\right.
\nonumber \\
&&+2(2q_{i+1,j,k}-q_{i+1,j+1,k}-q_{i+1,j-1,k})+(2q_{i+1,j,k}-q_{i,j,k}-q_{i+2,j,k})
\nonumber \\
&&+2(2q_{i,j-1,k}-q_{i,j-1,k+1}-q_{i,j-1,k-1})+(2q_{i,j-1,k}-q_{i,j,k}-q_{i,j-2,k})
\nonumber \\
&&+2(2q_{i,j+1,k}-q_{i,j+1,k+1}-q_{i,j+1,k-1})+(2q_{i,j+1,k}-q_{i,j,k}-q_{i,j+2,k})
\nonumber \\
&&+2(2q_{i,j,k-1}-q_{i+1,j,k-1}-q_{i,-1j,k-1})+(2q_{i,j,-1k}-q_{i,j,k}-q_{i,j-2,k})
\nonumber \\
&&\left.
+2(2q_{i,j,k+1}-q_{i+1,j,k+1}-q_{i-1,j,k+1})+(2q_{i,j,k+1}-q_{i,j,k}-q_{i,j,k+2})\right]
\nonumber \\
&&-\frac{p_A}2(N-1)\left[ q_{ijk}(t)-M(t)\right] +p_A(N-1)\frac{\left(
-K\right) }{2b}  \nonumber \\
&&\times \sum\limits_{\omega =\pm 1}\left[ \left( q_{i+\omega
,j,k}-M(t)\right) +\left( q_{i,j+\omega ,k}-M(t)\right) +\left(
q_{i,j,k+\omega }-M(t\right) \right] ,\ (b>K>0)  \label{7}
\end{eqnarray}
where $b>K>0$ is required by the convergence of the integration.

Now we determine the system behavior by studying the following order
parameters. First, from Eq. (\ref{6}) we obtain 
\begin{eqnarray}
\frac{d}{dt}M(t) &\equiv &\frac{1}{N}\sum\limits_{ijk}\frac{d}{dt}%
q_{ijk}(t)=p\frac{1}{N}\sum\limits_{ijk}Q_{ijk}^{G}+(1-p)\frac{1}{N}%
\sum\limits_{ijk}Q_{ijk}^{K}  \nonumber \\
&=&-p\left[ 1-\frac{6K}{b}-\frac{K}{b}(N-1)p_{A}\right] M(t),  \label{M}
\end{eqnarray}
$\sum_{ijk}Q_{ijk}^{K}=0$ means that the Kawasaki-type dynamics does not
change the value of $M\left( t\right) $. Second, we define 
\[
M^{\prime }(t)\equiv \frac{1}{N}\sum\limits_{ijk}q_{ijk}^{\prime }(t)\equiv 
\frac{1}{N}\sum\limits_{ijk}(-1)^{i+j+k}q_{ijk}(t), 
\]
and similarly we obtain 
\begin{equation}
\frac{d}{dt}M^{\prime }(t)\equiv \frac{1}{N}\sum\limits_{ijk}\frac{d}{dt}%
q_{ijk}^{\prime }(t)=p\frac{1}{N}\sum\limits_{ijk}Q_{ijk}^{\prime G}+(1-p)%
\frac{1}{N}\sum\limits_{ijk}Q_{ijk}^{\prime K},  \label{9}
\end{equation}
where 
\[
Q_{ijk}^{\prime G}=-q_{ijk}^{\prime }(t)-\frac{K}{b}\sum\limits_{\omega =\pm
1}\left[ q_{i+\omega ,j,k}^{\prime }(t)+q_{i,j+\omega ,k}^{\prime
}(t)+q_{i,j,k+\omega }^{\prime }(t)\right] +\frac{K}{b}%
(N-1)p_{A}(-1)^{i+j+k}M(t), 
\]
and 
\begin{eqnarray*}
Q_{ijk}^{\prime K} &=&\frac{1}{2(b-K)}b\left\{ \left[ (-q_{i+1,j,k}^{\prime
}-q_{i,j,k}^{\prime })-(q_{i,j,k}^{\prime }+q_{i-1,j,k}^{\prime })\right]
\right. \\
&&+\left[ (-q_{i,j+1,k}^{\prime }-q_{i,j,k}^{\prime })-(q_{i,j,k}^{\prime
}+q_{i,j-1,k}^{\prime })\right] +\left[ (-q_{i,j,k+1}^{\prime
}-q_{i,j,k}^{\prime })-(q_{i,j,k}^{\prime }+q_{i,j,k-1}^{\prime })\right] \}
\\
&&-\frac{K}{2(b-K)}\left[ 2(-2q_{i-1,j,k}^{\prime }-q_{i-1,j+1,k}^{\prime
}-q_{i-1,j-1,k}^{\prime })+(-2q_{i-1,j,k}^{\prime }-q_{i,j,k}^{\prime
}-q_{i-2,j,k}^{\prime })\right. \\
&&+2(-2q_{i+1,j,k}^{\prime }-q_{i+1,j+1,k}^{\prime }-q_{i+1,j-1,k}^{\prime
})+(-2q_{i+1,j,k}^{\prime }-q_{i,j,k}^{\prime }-q_{i+2,j,k}^{\prime }) \\
&&+2(-2q_{i,j-1,k}^{\prime }-q_{i,j-1,k+1}^{\prime }-q_{i,j-1,k-1}^{\prime
})+(-2q_{i,j-1,k}^{\prime }-q_{i,j,k}^{\prime }-q_{i,j-2,k}^{\prime }) \\
&&+2(-2q_{i,j+1,k}^{\prime }-q_{i,j+1,k+1}^{\prime }-q_{i,j+1,k-1}^{\prime
})+(-2q_{i,j+1,k}^{\prime }-q_{i,j,k}^{\prime }-q_{i,j+2,k}^{\prime }) \\
&&+2(-2q_{i,j,k-1}^{\prime }-q_{i+1,j,k-1}^{\prime
}-q_{i,-1j,k-1})+(-2q_{i,j,-1k}^{\prime }-q_{i,j,k}^{\prime
}-q_{i,j-2,k}^{\prime }) \\
&&\left. +2(-2q_{i,j,k+1}^{\prime }-q_{i+1,j,k+1}^{\prime
}-q_{i-1,j,k+1}^{\prime })+(-2q_{i,j,k+1}^{\prime }-q_{i,j,k}^{\prime
}-q_{i,j,k+2}^{\prime })\right] \\
&&-\frac{p_{A}(N-1)}{2}\left\{ q_{ijk}^{^{\prime
}}(t)-(-1)^{i+j+k}M(t)\right. \\
&&\left. -\frac{K}{b}\sum\limits_{\omega }(q_{i+\omega ,j,k}^{^{\prime
}}(t)+q_{i,j+\omega ,k}^{^{\prime }}(t)+q_{i,j,k+\omega }^{^{\prime }}(t))-%
\frac{6k}{b}(-1)^{i+j+k}M(t)\right\} .
\end{eqnarray*}
Performing the summation over the indices $i$, $j$, and $k$, we get 
\begin{equation}
\frac{d}{dt}M^{\prime }(t)=\left\{ -p\left( 1+\frac{6K}{b}\right)
+(1-p)\left[ 6\frac{6K-b}{b-K}-\frac{p_{A}(N-1)}{2}\left( 1-\frac{6K}{b}%
\right) \right] \right\} M^{\prime }(t).  \label{M'}
\end{equation}
The Solutions of equations (\ref{M}) and (\ref{M'}) are 
\begin{equation}
M(t)=M\left( 0\right) \exp \left\{ -p\left[ 1-\left( 1+\frac{(N-1)p_{A}}{6}%
\right) \frac{K}{K_{c}^{0}}\right] t\right\} ,\left( K>0\right)  \label{M-S}
\end{equation}
and 
\begin{eqnarray}
M^{\prime }(t) &=&M^{\prime }\left( 0\right) \exp \left\{ -\left( p\left( 1+%
\frac{K}{K_{c}^{0}}\right) +(1-p)\left[ \frac{36}{6-K/K_{c}^{0}}+\frac{%
p_{A}(N-1)}{2}\right] \left( 1-\frac{K}{K_{c}^{0}}\right) \right) t\right\} ,
\nonumber \\
&&\left( 
\begin{array}{c}
if\ p=1,then\ K>0 \\ 
if\ p\neq 1,then\ (b>K>0)
\end{array}
\right)  \label{M'-S}
\end{eqnarray}
where $K_{c}^{0}=\left\vert J\right\vert /k_{B}T_{c}=b/6$ is the critical
point of the three-dimensional Gaussian model without the SWN effect.

(1) When the SWN effect does not exist and $p_{A}=0$, 
\begin{equation}
M(t)=M\left( 0\right) \exp \left[ -p\left( 1-K/K_{c}^{0}\right) t\right] ,
\label{M-regular}
\end{equation}
\begin{equation}
M^{\prime }(t)=M^{\prime }\left( 0\right) \exp \left\{ -\left[ p\left(
1+K/K_{c}^{0}\right) +36(1-p)\frac{1-K/K_{c}^{0}}{6-K/K_{c}^{0}}\right]
t\right\} .  \label{M'-regular}
\end{equation}
Analyzing the long-time system behavior, we find that:

(1.a) For the case of 
\[
K<K_{c}^{0},\text{ }\quad (T>T_{c}^{0}), 
\]
we have both vanishing $M\left( t\right) $ and $M^{\prime }\left( t\right) $%
, which correspond to the paramagnetic phase.

(1.b) For the case of 
\begin{equation}
K\rightarrow K_{c}^{0}
\end{equation}
$M^{\prime }\left( t\right) \rightarrow 0$, and 
\begin{equation}
M(t)=M\left( 0\right) \exp \left[ -p\frac{t}{\tau }\right] ,\tau =\frac{1}{%
1-K/K_{c}^{0}}\rightarrow \infty .
\end{equation}
The critical slowing down of the order parameter $M(t)$ will appear at the
critical point $K_{c}^{0}$.

(1.c) For the case of 
\[
K>K_{c}^{0},\text{ }\quad (T<T_{c}), 
\]
and 
\[
\frac{K}{K_{c}^{0}}<\frac{1}{2p}\left( -36+41p+\sqrt{\left(
1561p^{2}-2808p+1296\right) }\right) , 
\]
we have nonvanishing $M\left( t\right) $ and vanishing $M^{\prime }\left(
t\right) $, which correspond to the ferromagnetic phase.

(1.d) For the case of 
\[
\frac{1}{2p}\left( -36+41p+\sqrt{\left( 1561p^{2}-2808p+1296\right) }\right)
<\frac{K}{K_{c}^{0}}<6 
\]
we have both nonvanishing $M\left( t\right) $ and $M^{\prime }\left(
t\right) $.

With both of the order parameters nonvanishing, this phase cannot be simply
identified as ferromagnetic or antiferromagnetic. We name it as heterophase,
and from Eqs. (\ref{M-regular}) and (\ref{M'-regular}) we can see that the
system behavior strongly depends on the initial condition and the
temperature.

(1.e) For the case of 
\[
K/K_{c}^{0}\geq 6, 
\]
we could obtain $M\left( t\right) \neq 0$, and when $p=1$ (with pure
Glauber-type dynamics simulating the contact with a heat bath), we have $%
M^{\prime }\left( t\right) \rightarrow 0$. However, with $p\neq 1$ (with an
external energy flow), because the antiferromagnetic Kawasaki-type dynamics
is limited by the condition ($b>K>0$) to ensure the convergence of the
integration, we cannot obtain the value of $M^{\prime }\left( t\right) $. As
a result, we cannot theoretically obtain the system behavior in the
temperature region $K/K_{c}^{0}\geq 6$\footnote{%
The same problem exists in Ref. \cite{zhu3}, i.e., the system behavior
cannot be theoretically obtained for the region $K\geq b.$}.

The phase diagram is shown in Fig. 1(a).

(2) Now the A-SWN effect is introduced and $p_A\neq 0$. We suppose $p_A=1/N$%
, and then we have

\begin{equation}
M(t)=M\left( 0\right) \exp \left\{ -p\left[ 1-\left( 1+\frac{1}{6}\right) 
\frac{K}{K_{c}^{0}}\right] t\right\} ,
\end{equation}
and 
\begin{equation}
M^{\prime }(t)=M^{\prime }\left( 0\right) \exp \left\{ -\left( p\left(
1+K/K_{c}\right) +(1-p)\left[ \frac{36}{6-K/K_{c}^{0}}+\frac{1}{2}\right]
\left( 1-\frac{K}{K_{c}^{0}}\right) \right) t\right\} .
\end{equation}
Analyzing the long-time asymptotic behavior, we can similarly get:

(2.a) For the case of 
\[
K<\left. K_{c}\right| _{p_{A}=1/N}=\frac{6}{7}K_{c}^{0}, 
\]
the condition leads to vanishing $M\left( t\right) $ and vanishing $%
M^{\prime }\left( t\right) $, which correspond to the paramagnetic phase.

(2.b) For the case of 
\[
K\rightarrow \left. K_{c}\right\vert _{p_{A}=1/N}=\frac{6}{7}K_{c}^{0}, 
\]
The critical slowing down of the system will appear.

(2.c) For the case of 
\[
\frac{89p-79+\sqrt{7129p^{2}-12862p+5929}}{2\left( 3p-1\right) }>\frac{K}{%
K_{c}^{0}}>\frac{6}{7} 
\]
we have nonvanishing $M\left( t\right) $ and vanishing $M^{\prime }\left(
t\right) $, which correspond to the ferromagnetic phase.

(2.d) For the case of 
\[
6>\frac{K}{K_{c}^{0}}>\frac{89p-79+\sqrt{7129p^{2}-12862p+5929}}{2\left(
3p-1\right) }, 
\]
we have both nonvanishing $M\left( t\right) $ and $M^{\prime }\left(
t\right) $, which correspond to the heterophase.

(2.e) For the case of 
\[
K/K_{c}^{0}\geq 6, 
\]
For the same reason (see (1.e)), we cannot obtain theoretically the system
behavior.

The phase diagram is shown in Fig. 1(b).

\section{Competing mechanism on R-SWN}

\label{Sec.3}

Now we investigate the competing dynamics on rewiring-type small-world
networks (the definition see the Introduction), and the sequence of the
content is the same as that in the previous section.

With the Glauber-type mechanism, the master equation has the same form as
Eq. (\ref{3}), and the single-spin evolving equation is also given by Eq. (%
\ref{q-Glauber}).

With the Kawasaki-type mechanism, the master equation can be written as 
\begin{eqnarray*}
\frac d{dt}P(\{\sigma \},t) &=&(1-p_R)\sum\limits_{\langle jl\rangle
}\sum\limits_{\hat{\sigma}_j\hat{\sigma}_l}\left[ -W_{jl}(\sigma _j\sigma
_l\rightarrow \hat{\sigma}_j\hat{\sigma}_l)P(\{\sigma \},t)\right. \\
&&\left. +W_{jl}(\hat{\sigma}_j\hat{\sigma}_l\rightarrow \sigma _j\sigma
_l)P(\{\sigma _{j\neq i},\sigma _{l\neq k}\},\hat{\sigma}_{j,}\hat{\sigma}%
_l,t)\right] \\
&&+\frac{Dp_R}{N-1}\sum\limits_j\sum\limits_{l\neq j}\sum\limits_{\hat{\sigma%
}_j\hat{\sigma}_l}\left[ -W_{jl}(\sigma _j\sigma _l\rightarrow \hat{\sigma}_j%
\hat{\sigma}_l)P(\{\sigma \},t)\right. \\
&&\left. +W_{jl}(\hat{\sigma}_j\hat{\sigma}_l\rightarrow \sigma _j\sigma
_l)P(\{\sigma _{j\neq i},\sigma _{l\neq k}\},\hat{\sigma}_{j,}\hat{\sigma}%
_l,t)\right] .
\end{eqnarray*}
The redistribution probability $W_{jl}$ is given by Eq. (\ref{5}), and the
single-spin evolving equation 
\begin{eqnarray*}
\frac d{dt}q_k(t) &=&Q_k^K=(1-p_R)\left\{ -2Dq_k(t)+\sum\limits_{\{\sigma
\}}\sum\limits_{\omega =\pm 1}\left[ \sum\limits_{\hat{\sigma}_k,\hat{\sigma}%
_{k+\omega }}\hat{\sigma}_kW_{k,k+\omega }(\sigma _k\sigma _{k+w}\rightarrow 
\hat{\sigma}_k\hat{\sigma}_{k+w})\right] P(\{\sigma \},t)\right\} \\
&&+\frac{Dp_R}{N-1}\left\{ -(N-1)q_k(t)+\sum\limits_{\{\sigma \}}\left[
\sum\limits_{l\neq k}\sum\limits_{\hat{\sigma}_k,\hat{\sigma}_l}\hat{\sigma}%
_kW_{kl}(\sigma _k\sigma _l\rightarrow \hat{\sigma}_k\hat{\sigma}_l)\right]
P(\{\sigma \},t)\right\} .
\end{eqnarray*}

With the competing mechanism, the evolution of $q_k(t)$ is once again
described by an equation having the same form as Eq. (\ref{6}) 
\[
\frac d{dt}q_k(t)=pQ_k^G+(1-p)Q_k^K, 
\]
which consists of two terms, one corresponding to the Glauber-type dynamics
and the other corresponding to the Kawasaki-type dynamics. In the following
we turn to the Gaussian model.

For the three-dimensional Gaussian model built on a R-SWN, the Hamiltonian
can be written as 
\begin{eqnarray}
-\beta H &=&K(1-p_R)\sum\limits_{ijk}\sigma _{ijk}\left( \sigma
_{i+1,j,k}+\sigma _{i,j+1,k}+\sigma _{i,j,k+1}\right)  \nonumber \\
&&+\frac 3Np_RK\sum\limits_{ijk}\sigma _{ijk}\sum\limits_{i^{\prime
}j^{\prime }k^{\prime }}\sigma _{i^{\prime }j^{\prime }k^{\prime }}-\frac 3N%
p_RK\sum\limits_{ijk}\sigma _{ijk}^2.  \label{R-SWN}
\end{eqnarray}
In the single-spin evolving equation with the competing mechanism, the term
that corresponds to the Glauber-type dynamics is now given by 
\[
Q_{ijk}^G=-q_{ijk}(t)+\frac Kb(1-p_R)\sum\limits_{\omega =\pm 1}\left(
q_{i+\omega ,j,k}+q_{i,j+\omega ,k}+q_{i,j,k+\omega }\right) +\frac{6p_R}b%
KM(t),\ \left( K>0\right) 
\]
and the term that corresponds to the Kawasaki-type dynamics can be written
as (similarly, we switch $\beta $ to -$\beta $, and $K$ to $-K$) 
\begin{eqnarray*}
Q_{ijk}^K &=&\frac 1{2[b-K(1-p_R)]}b\left\{ \left[
(q_{i+1,j,k}-q_{i,j,k})-(q_{i,j,k}-q_{i-1,j,k})\right] \right. \\
&&+\left. \left[ (q_{i,j+1,k}-q_{i,j,k})-(q_{i,j,k}-q_{i,j-1,k})\right]
+\left[ (q_{i,j,k+1}-q_{i,j,k})-(q_{i,j,k}-q_{i,j,k-1})\right] \right\} \\
&&-\frac{K(1-p_R)}{2\left[ b-K(1-p_R)\right] }\left[ 2\left(
2q_{i-1,j,k}-q_{i-1,j+1,k}-q_{i-1,j-1,k}\right) +\left(
2q_{i-1,j,k}-q_{i,j,k}-q_{i-2,j,k}\right) \right. \\
&&+2\left( 2q_{i+1,j,k}-q_{i+1,j+1,k}-q_{i+1,j-1,k}\right) +\left(
2q_{i+1,j,k}-q_{i,j,k}-q_{i+2,j,k}\right) \\
&&+2\left( 2q_{i,j-1,k}-q_{i,j-1,k+1}-q_{i,j-1,k-1}\right) +\left(
2q_{i,j-1,k}-q_{i,j,k}-q_{i,j-2,k}\right) \\
&&+2\left( 2q_{i,j+1,k}-q_{i,j+1,k+1}-q_{i,j+1,k-1}\right) +\left(
2q_{i,j+1,k}-q_{i,j,k}-q_{i,j+2,k}\right) \\
&&+2\left( 2q_{i,j,k-1}-q_{i+1,j,k-1}-q_{i,-1j,k-1}\right) +\left(
2q_{i,j,-1k}-q_{i,j,k}-q_{i,j-2,k}\right) \\
&&+2\left( 2q_{i,j,k+1}-q_{i+1,j,k+1}-q_{i-1,j,k+1}\right) +\left(
2q_{i,j,k+1}-q_{i,j,k}-q_{i,j,k+2}\right) \\
&&-\frac 32p_R\left\{ q_{ijk}(t)-M(t)+\frac{K(1-p_R)}b\left[
\sum\limits_{\omega =\pm 1}\left( q_{i+\omega ,j,k}+q_{i,j+\omega
,k}+q_{i,j,k+\omega }\right) -6M\left( t\right) \right] \right\} , \\
&&\left( b>K(1-p_R)>0\right)
\end{eqnarray*}
where $b>K(1-p_R)>0$ is required by the convergency of the integration.

From the single-spin evolving equation with the competing mechanism, we can
obtain 
\begin{equation}
\frac d{dt}M(t)=\frac 1N\sum\limits_{ijk}\frac d{dt}q_{ijk}(t)=p\left( -1+%
\frac{6K}b\right) M(t),  \label{11}
\end{equation}
and 
\begin{eqnarray}
\frac d{dt}M^{\prime }(t) &=&p\frac 1N\sum\limits_{ijK}Q_{ijk}^{\prime
G}+(1-p)\frac 1N\sum\limits_{ijk}Q_{ijk}^{\prime K}  \nonumber \\
&=&\left( -p\left[ 1+\frac{6K}b\left( 1-p_R\right) \right] \right.  \nonumber
\\
&&\left. +(1-p)\left\{ 6\frac{6K\left( 1-p_R\right) -b}{b-K\left(
1-p_R\right) }-\frac{3p_R}2\left[ 1-\frac{6K}b\left( 1-p_R\right) \right]
\right\} \right) M^{\prime }(t).  \label{12_1}
\end{eqnarray}
The solutions of Eqs. (\ref{11}) and (\ref{12_1}) are, respectively, 
\begin{equation}
M(t)=M\left( 0\right) \exp \left[ -p(1-K/K_c^0)t\right] ,\ \left( K>0\right)
\label{13}
\end{equation}
and 
\begin{eqnarray}
M^{\prime }(t) &=&M^{\prime }\left( 0\right) \exp \left( -\left\{ p\left[
1+(1-p_R)\frac K{K_c^0}\right] \right. \right.  \nonumber \\
&&\left. \left. +(1-p)\left[ \frac{36}{6-K(1-p_R)/K_c^0}+\frac{3p_R}2\right]
\left[ 1-(1-p_R)\frac K{K_c^0}\right] \right\} \right) ,  \nonumber \\
&&\left( 
\begin{array}{c}
if\ p=1,then\ K>0 \\ 
if\ p\neq 1,then\ \left( b>K(1-p_R)>0\right)
\end{array}
\right) ,  \label{14}
\end{eqnarray}
where $K_c^0=J/k_BT_c^0=b/6$ is the critical point without the SWN effect.

(1) When no SWN effect is considered and $p_R=0$, the phase diagram is given
by Fig. 1 (a).

(2) Now we introduce the R-SWN effect and set $p_{R}=0.1$, and then Eqs. (%
\ref{13}) and (\ref{14}) become 
\begin{equation}
M(t)=M\left( 0\right) \exp \left[ -p(1-K/K_{c}^{0})t\right] ,
\label{M-solution-RSWN}
\end{equation}
and 
\begin{equation}
M^{\prime }(t)=M^{\prime }\left( 0\right) \exp \left\{ -\left[ p\left( 1+%
\frac{9K}{10K_{c}^{0}}\right) +(1-p)\left( \frac{36}{6-\frac{9K}{10K_{c}^{0}}%
}+\frac{3}{20}\right) \left( 1-\frac{9K}{10K_{c}^{0}}\right) \right]
\right\} .  \label{15}
\end{equation}
Analyzing the long-time asymptotic behavior, we find that:

(2.a) For the case of 
\[
K<K_{c}^{0},\text{ }\left( T>T_{c}^{0}\right) , 
\]
we have both vanishing $M\left( t\right) $ and $M^{\prime }\left( t\right) $%
, and this corresponds to the paramagnetic phase.

(2.b) For the case of 
\[
K\rightarrow K_{c}^{0}, 
\]
we shall observe the critical slowing down of the system.

(2.c) For the case of 
\[
K>K_{c}^{0},\text{ }(T<T_{c}^{0}), 
\]
and

\[
\frac K{K_c^0}<\frac{25230p-22230+150\sqrt{26017p^2-46842p+21609}}{2\left(
621p-81\right) }, 
\]
the condition leads to nonvanishing $M\left( t\right) $ and vanishing $%
M^{\prime }\left( t\right) $, which correspond to the ferromagnetic phase.

(2.d) For the case of 
\[
\frac{25230p-22230+150\sqrt{26017p^{2}-46842p+21609}}{2\left( 621p-81\right) 
}<\frac{K}{K_{c}^{0}}<\frac{20}{3}, 
\]
we have both nonvanishing $M\left( t\right) $ and $M^{\prime }\left(
t\right) $, which correspond to the heterophase.

(2.e) For the case of 
\[
K/K_c^0\geq 20/3, 
\]
For the same reason (see Sec. II. (1.e)), we cannot give theoretically the
behavior of the system in this region.

The phase diagram is shown in Fig. 1(c).

\section{Summary}

\label{Sec.4}

Comparing Fig. 1(a), (b) and (c), we find both similarities and differences.
The phase diagrams have similar structures while the boundaries may be
shifted by the SWN effect. In the following we briefly explain our
observations and discuss the nature of the competing mechanism and the SWN
effect.

First we notice that the phase diagrams are all separated into four regions,
with two special temperatures: one is the critical temperature $K_c=J/k_BT_c$%
, another is the limit temperature $K_{\max }=J/k_BT_{\min }$\ required by
the convergence of the integration. When the SWN effect does not exist, $%
K_c=K_c^0=J/k_BT_c^0=b/2D$ is the critical point of the Gaussian model,
where $b$\ is the Gaussian distribution constant and $D$\ is the space
dimension, and $K_{\max }=b$\ (since $b>K=\left| J\right| /k_BT>0${\bf \ }is
the region where we can assure the convergence of the integration in the
Kawasaki-type dynamics of the antiferromagnetic system). When the A-SWN
effect is introduced, $K_c=\frac 6{6+(N-1)p_A}K_c^0$ and $K_{\max
}=b=2DK_c^0 $ (in the example studied $D=3$, $p_A\approx 1/N$, and $%
K_c/K_c^0=6/7$).{\bf \ }When the R-SWN effect is introduced, $K_c=K_c^0$\
and $K_{\max }=\frac b{(1-p_R)}=\frac{2DK_c^0}{(1-p_R)}$\ (in the example
studied $D=3$, $p_R=0.1$, and $K_{\max }/K_c^0=20/3$). Now we describe the
phase behavior: Above the critical temperature $T_c=J/k_BK_c$, one can only
observe a disordered state, namely the paramagnetic phase, because of the
dominating heat noise. Below critical temperature $T_c$, the system begins
to show some kind of order, and the system behavior is determined by the
competition between the Glauber-type mechanism which favors a ferromagnetic
state, and the Kawasaki-type mechanism which favors an antiferromagnetic
state. The result of the competition is subject to two factors, the
probability of each mechanism and the temperature. However, the
Kawasaki-type dynamics describing an antiferromagnetic Gaussian system is
limited by the condition $T>T_{\min }$, which is necessary for the
convergence of the integration. As a result, in the region $T\leq T_{\min }$
we cannot theoretically obtain the system behavior.

Now we discuss how and why the two special temperatures $T_{c}$\ and $%
T_{\min }$\ may be affected by the SWN effect. When the Gaussian model is
built on a R-SWN, the temperature $T_{c}\left( =T_{c}^{0}\right) $ remains
the same. This is because, as is clear in all of the phase diagrams, $T_{c}$
is determined by the competition between the heat noise and the Glauber-type
mechanism, which favors a lower energy state. One important characteristic
of the Gaussian model is that this temperature can be further determined by
the average coordination number. For example, without the SWN effect this
temperature is given by $J/k_{B}T_{c}=b/2D$, where $2D$ is the average
coordination number of a $D$-dimensional lattice. On a R-SWN, the average
coordination number is not changed when a portion of the regular links are
rewired, and thus $T_{c}$ also remains unchanged. On the other hand the
temperature $T_{\min }$ is lower than the value on a regular network. On
R-SWN long range links have made long range spin-pair redistribution
possible, and therefore the influence of the Kawasaki-type mechanism is
enhanced. Thus it is not difficult to understand why on R-SWN there is a
lower value of the temperature $T_{\min }$.

When the Gaussian model is built on an A-SWN, the temperature $T_{c}\left( =%
\frac{6+(N-1)p_{A}}{6}T_{c}^{0}\right) $ becomes higher. As is explained
above, this temperature is determined by the average coordination number,
which is $\frac{7}{6}\cdot 2D$ on a three dimensional A-SWN with $%
p_{A}\approx 1/N$. At the same time the temperature $T_{\min }$ is
unchanged. We may explain that it is because the two dynamics are both
strengthened, i.e., by the larger coordination number and the long range
spin-pair redistribution, and the temperature that characterizes a certain
counterbalance remains the same.

To summarize, in this article we investigate the competing Glauber-type and
Kawasaki-type dynamics on two typical three-dimensional small-world
networks, adding-type (A-SWN) and rewiring -type (R-SWN), in the framework
of the Gaussian model. We get the evolution of the order parameters, $M(t)$
and $M^{\prime }(t)$, and by analyzing the long time asymptotic behavior we
draw the phase diagrams. With the competing mechanism, there exist two
easily identifiable special temperatures. The influence of the long range
links is analyzed: For the Glauber-type dynamics, more important is the
altered average coordination number while the Kawasaki-type dynamics is
enhanced by the long range spin interaction and redistribution.

\acknowledgments

This work was supported by the National Natural Science Foundation of China
under Grant No. 10075025.

\null\vskip0.2cm

\centerline{\bf Caption of figures} \vskip1cm

Fig. 1 The phase diagrams of the three-dimensional Gaussian model on (a)
regular lattice, (b) A-SWN, and (c) R-SWN. The regions Para, Ferro, and
Hetero correspond to the paramagnetic, ferromagnetic, and heterophase phase,
respectively, while the part of oblique lines belong to unknown region.

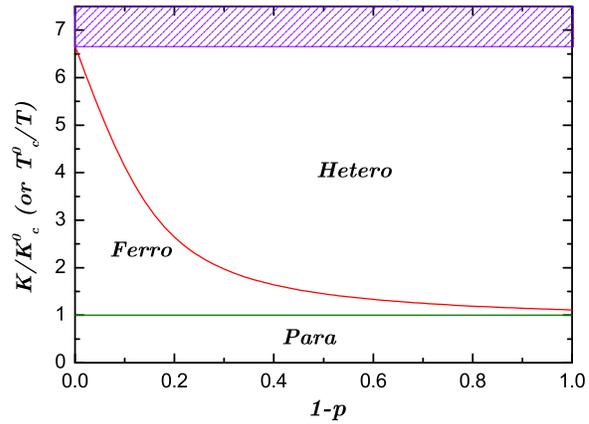
\end{document}